\documentclass[twocolumn,floatfix,showpacs,prl,superscriptaddress]{revtex4-1}
\usepackage{graphicx}
\usepackage{amsmath}
\usepackage{amssymb}
\usepackage{bm}
\usepackage{color}

\begin{document}

\title{Magnetic impurities on the surface of topological superconductor}
\author{Ting-Pong Choy}
\affiliation{Department of Physics, Hong Kong University of Science and Technology, Clear Water Bay Road, Hong Kong, China}
\affiliation{Institute for Advanced Study, Hong Kong University of Science and Technology, Hong Kong, China}
\author{K. T. Law}
\affiliation{Department of Physics, Hong Kong University of Science and Technology, Clear Water Bay Road, Hong Kong, China}
\author{Tai-Kai Ng}
\affiliation{Department of Physics, Hong Kong University of Science and Technology, Clear Water Bay Road, Hong Kong, China}

\begin{abstract}
We consider the effects of magnetic impurities on the surface of superconducting Cu doped $Bi_{2}Se_{3}$ in the odd parity pairing phase which support topologically protected Majorana fermions surface states with linear spectrum. We show that a single magnetic impurity on the surface may induce a pair of in-gap localized bound states. The energy of the in-gap state is extremely sensitive to the orientation of the magnetic impurity due to the so-called Ising properties of Majorana fermions. The magnetic impurity induced spin-texture, which can be measured using spin sensitive STM, is calculated. We also show that the RKKY interactions between magnetic impurities mediated via the Majorana fermions are always ferromagnetic and dense enough magnetic impurities will develop long-range magnetic order and break the time-reversal symmetry on the material surface eventually.
\end{abstract}

\maketitle

{\bf \emph{Introduction}}--- Topological superconductors (TSCs) are new states of matter with bulk pairing gap and symmetry protected gapless boundary modes.\cite{alicea10,oreg,fukane08,potter,lee09,duckheim,weng11} The zero energy surface Andreev bound states (SABSs) are Majorana fermions (MFs) as they act as their own anti-particles due to particle-hole (PH) symmetry. It has been shown that the self-Hermitian property of MFs is the origin of several interesting phenomena such as the MF induced resonant Andreev reflection\cite{law09,wimmer} and the $4\pi$ Josephson effect \cite{kitaev,kwon,fu09,lutchyn,law11,heck}. Particularly, it has been shown that MFs boundary states only interact with one specific direction the magnetic impurities where the specific direction is determined by the pairing symmetry of the bulk superconductor and the surface orientation. This is called the Ising spin properties of MFs. The consequences of the Ising spin properties of MFs for the surface states of He3 B phase\cite{chung2009} and for edge states of two-dimensional superconductors\cite{shindou} have been studied. It is shown that the magnetic susceptibilities of magnetic impurities which interact with MFs exhibit strong anisotropy due to the Ising properties of MFs. However, the Ising spin properties of MFs of a three dimensional TSC are yet to be explored.

The recently discovered superconducting topological insulator (STI) Cu$_x$Bi$_2$Se$_3$ is a possible realization of a three dimensional TSC.\cite{hor} It has been proposed that Cu$_x$Bi$_2$Se$_3$ is a TSC with odd-parity triplet pairing and fully gapped in the bulk \cite{fu2010} and this superconducting phase support SABSs with linear dispersion.  Specific heat \cite{kriener} and point contact spectroscopy measurements\cite{sasaki,kirzhner} of the material are consistent with this odd-parity triplet pairing phase.

In this work, we consider the effects of magnetic impurities coupled to the MFs emerged on the surface of an odd-parity triplet pairing TSC. We show that the magnetic impurity induced in-gap bounded states are extremely sensitive to the orientations of the magnetic moment due to the Ising properties of MFs. The spin texture near a magnetic impurity, which can be measured using spin-resolved STM, is calculated. Moreover, we show that the RKKY interactions between magnetic impurities mediated by the Ising MFs are always ferromagnetic. Finite density of magnetic impurities can break time-reversal symmetry spontaneously and open an energy gap on the surface. When this happens, the TS becomes a quantum thermal Hall insulator. 

\begin{figure}[tb]
\begin{tabular}{cc}
\includegraphics[width=0.3\linewidth,angle=90]{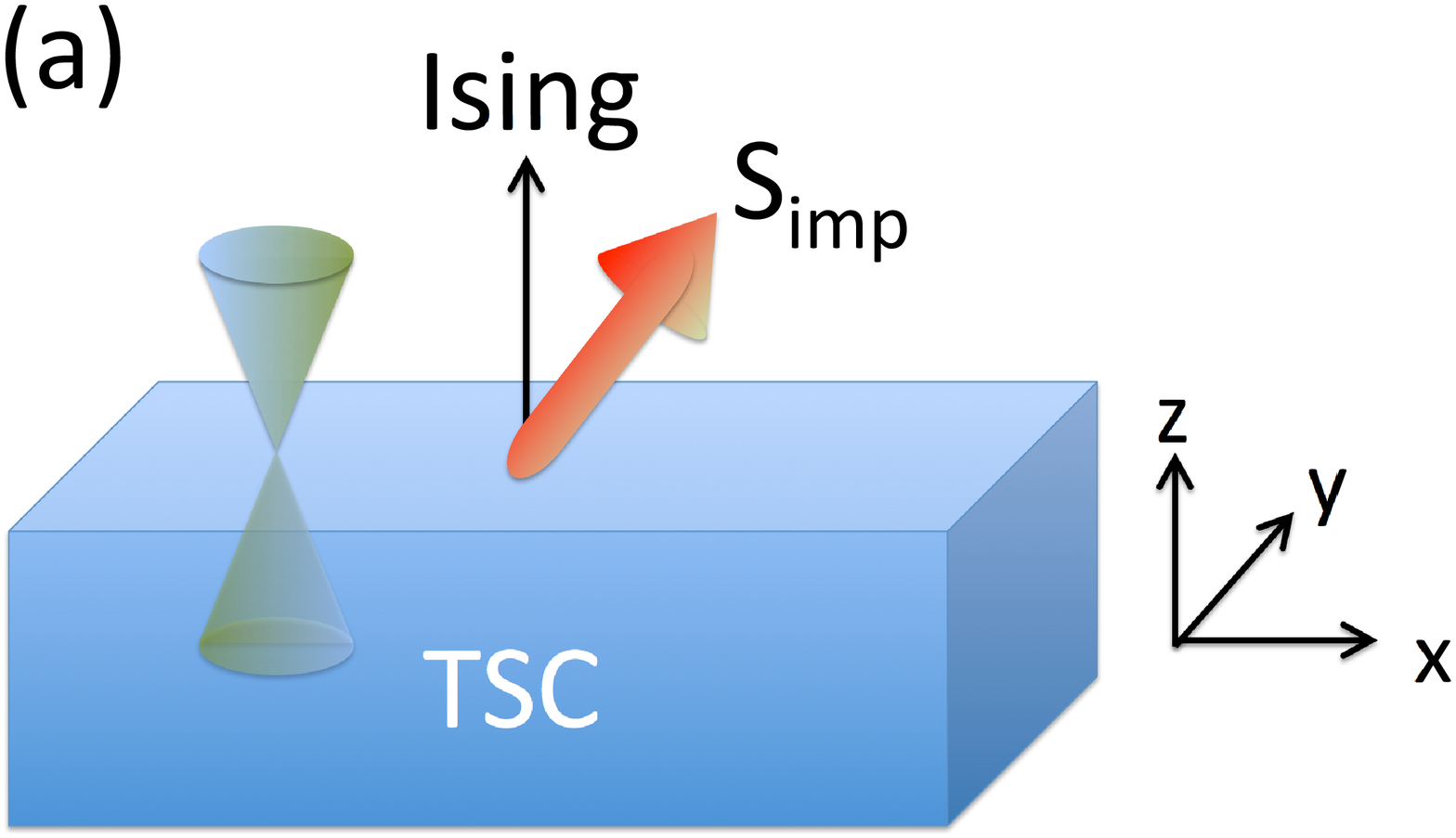}&
\includegraphics[width=0.48\linewidth]{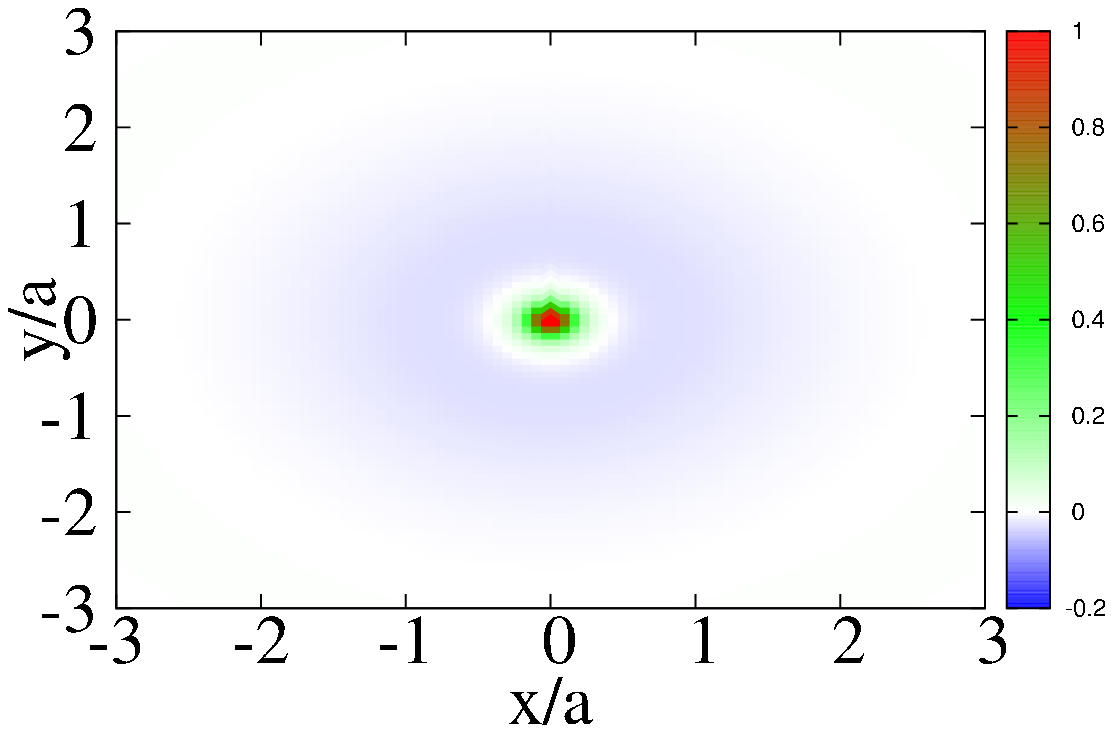}\\
\includegraphics[width=0.48\linewidth]{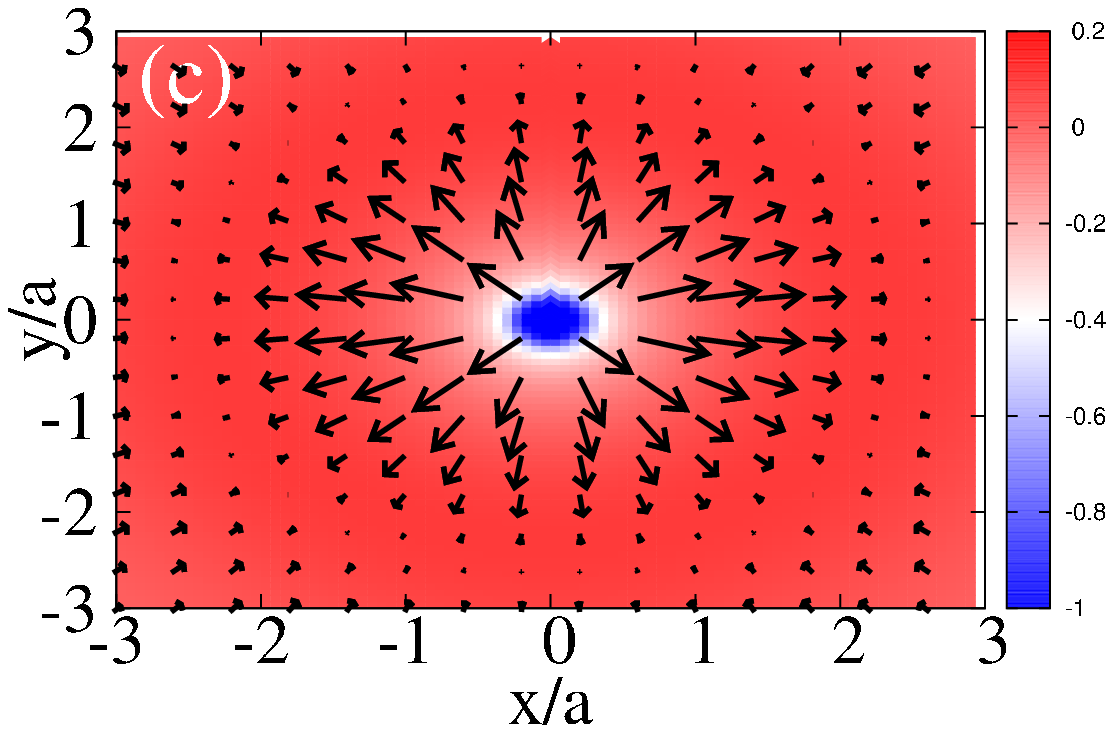}&
\includegraphics[width=0.48\linewidth]{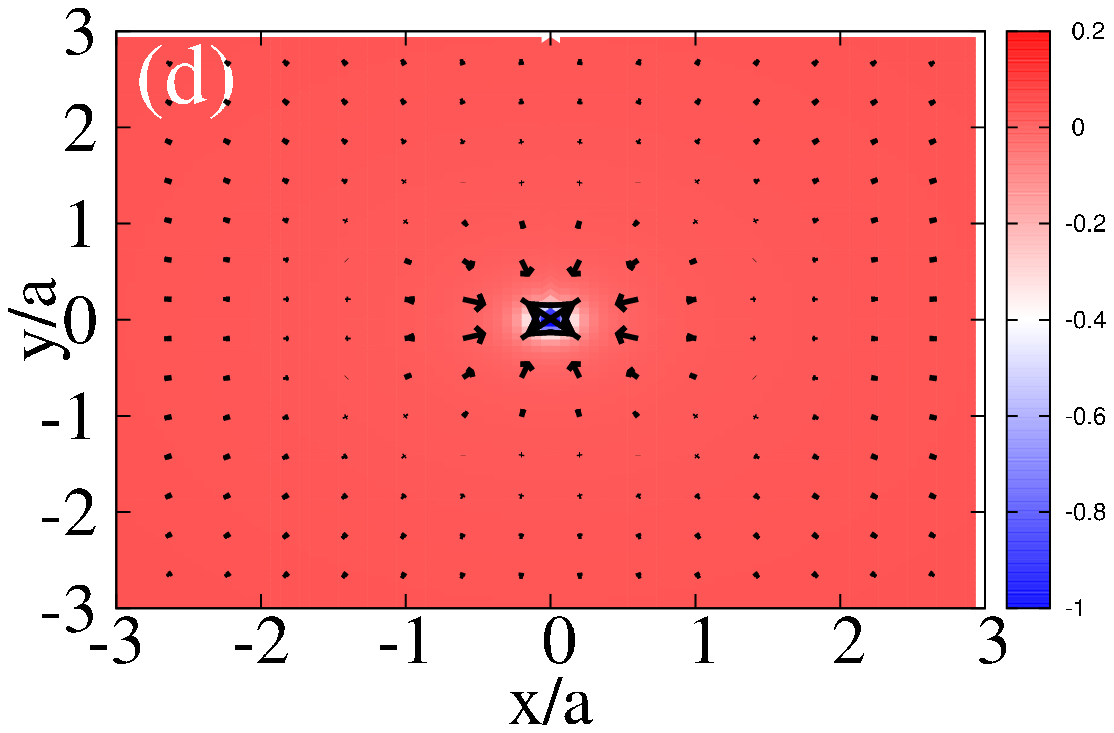}
\end{tabular}
\caption{\label{setup} (a) Illustration of a TSC with a single Majorana cone on the surface coupled to a magnetic impurity. The Ising spin direction of the Majorana modes is normal to the surface. (b) The modulation of the LDOS $\rho(\mathbf R, -0.7\Delta)$ around a magnetic impurity with scattering strength $U/\Delta=1$ which is polarized along the z-axis ($n_z = 1.0$) is shown. The corresponding pin textures at energy $\omega = -0.7\Delta$ (c) and $\omega=-0.5\Delta$ (d) near the magnetic impurity are shown. The component in the xy plane is denoted by a vector while the background colour indicates the magnitude of $s_z(\mathbf r,\omega)$. The arrows are normalized to the longest  in-plane spin length and $a=\hbar v_s/\Delta$.} 
\end{figure}

{\bf \emph{MF Ising spin}}--- We start with the Bogoliubov-de Gennes (BdG) Hamiltonian of  Cu$_x$Bi$_2$Se$_3$ given by\cite{fu2010}:
\begin{align}
\label{H0}
H &= \int d^3 \mathbf{k} \xi^\dagger_k H(\mathbf k) \xi_k,\\
H(k) &= [H_0(k)-\mu]\tau_z + \Delta \sigma_y s_z \tau_x.
\end{align}
Here $\xi_k = (c_{k\uparrow}, c_{k\downarrow}, c^\dagger_{-k\downarrow}, -c^\dagger_{-k\uparrow})^T$  are Nambu basis denoted by the Pauli matrices $\tau_{x,y,z}$ and $\uparrow,\downarrow$ are the electron spin indices. $s_{x,y,z}$ are Pauli matrices. At low energy, the band structure of the parent compound Bi$_2$Se$_3$ is well described by the $k\cdot p$ Hamiltonian:
\begin{equation} \label{BdG}
H_0(\mathbf k) = m\sigma_x + v(k_x \sigma_z s_y - k_y \sigma_z s_x) + v_z k_z \sigma_y,
\end{equation}
where $\sigma_z = \pm 1$ is an orbital index which denotes the two Se p$_z$ orbitals on the top and bottom layer in each unit cell. When $m=0$, the bulk is a single species of noninteracting three-dimensional Dirac quasiparticles. 

The  BdG Hamiltonian describes a DIII class \cite{schnyder,kitaev} superconductor which satisfies $\Theta H(\mathbf k) \Theta^{-1} = H(-\mathbf k)$ and $J H(\mathbf k) J^{-1} = - H(-\mathbf k)$, where $\Theta = i s_y K$ and $J = s_y \tau_y K$ are the time-reversal symmetry and particle-hole symmetry operators respectively. Interestingly, the Hamiltonian satisfy an extra inversion ($Z_2$) symmetry $\sigma_x\tau_z H(\mathbf k)\sigma_x\tau_z = H(-\mathbf k)$
where $\sigma_x$ is an inversion operator interchanging two orbitals. In this work, we consider the superconducting pairing denoted by $\hat\Delta = \Delta \sigma_y s_z \tau_x$ with $\sigma_x \hat\Delta \sigma_x = - \hat\Delta$ which is an odd-parity inter-orbital triplet pairing. Such pairing symmetry is consistent with the crystal point group $D_{3d}$ of the Cu$_x$Bi$_2$Se$_3$ and fully gapped in the bulk. A superconductor described by Eq.\ref{BdG} can support topologically protected MF surface states as shown below .\cite{fu2010}

To study the surface states, we consider the case where the superconductor is terminated at the $z=0$ plane where the wave function on the topmost layer vanishes, i.e. $\sigma_z \psi |_{z=0} = - \psi |_{z=0}$. By solving the semi-infinite BdG equation at $k_x = k_y =0$, we find a Kramers pair of zero-energy surface Andreev bound states $\psi_{\pm}$,
\begin{align}
\psi_{s} (z) &= A e^{\kappa z} 
\left( 
\begin{array}{c}
\sin(k_Fz)\\
\sin(k_Fz+\theta)
\end{array}
\right)_\sigma
\otimes |s_z = s, \tau_y = {\rm sgn}(v_z) s\rangle,
\end{align}
where $\kappa = \Delta/|v_z|$, $v_z k_F = \sqrt{\mu^2-m^2}$ and $e^{i\theta} = (m + i\sqrt{\mu^2-m^2})/\mu$. $A$ is the normalziation constant for the wavefunctions. By $\mathbf k\cdot \mathbf p$ theory, we obtain the low-energy 
Hamiltonian $H_s$ for surface Andreev bound states by expanding at small momentum,
\begin{align}
\label{hs}
H_s (\mathbf k) = v_s \left( k_x \tilde s_y - k_y \tilde s_x \right),
\end{align}
where the renormalized velocity in the effective Hamiltonian is,
\begin{align}
\frac{v_s}{v} = \frac{ \kappa( 1 - \cos 2\theta ) + k_F \sin 2\theta}
{\frac{2}{\kappa}(\kappa^2+k_F^2)
- \kappa(1 + \cos 2\theta ) + k_F \sin 2\theta}.
\end{align}
The effective Hamiltonian describes the gapless Majorana fermions on the surface boundary up to a high energy cutoff $\Delta$ within the bulk gap. In general, $\tilde s = (\tilde s_x, \tilde s_y, \tilde s_z)$ are SU(2) Pauli matrices which describes the coupling between two branches $\psi_s(\mathbf k, z)$ with opposite $s_z$ and $\tau_y$. In this explicit model they are  identical to the physical spin, $(s_x,s_y,s_z)$. The velocity of the boundary Majorana modes has a sign change depending on the value of $m$ and $v_s \sim v\Delta^2/\mu^2$ as $m\rightarrow 0$. The sign change corresponds to a structural transition of energy dispersion of the surface modes.\cite{yamakage} For positive value of $m$, the energy spectrum of SABS forms a Dirac cone. Indeed, when $m$ becomes negative, a second crossing of the zero energy appears at finite k and this crossing is protected by the band-inversion of the parent topological insulator.\cite{hsieh}

The eigenstates of the effective Hamiltonian $H_s$ form two branches $\phi_{\pm}(\mathbf k) = \frac{1}{\sqrt{2}} (1,\pm i 
e^{i\theta_k})^T$ with energy $E_k = \pm v_s k$  which are localized on the boundary $z=0$. Because of the PH symmetry, 
two branches are not independent and satisfy: $\phi_+(\mathbf k) = \phi_-(-\mathbf k)$.

To study the interaction between the MF surface states and a local magnetic impurity, we need to construct the local electron operators. By rotating the spin quantization axis along the x-direction $\psi_\rightarrow(\mathbf r), \psi_\leftarrow(\mathbf r)$, the surface mode expansion of the local electron operators can be written as:
\begin{align}
\left( 
\begin{array}{c}
\psi_\rightarrow(\mathbf r)\\
\psi_\leftarrow(\mathbf r)\\
\psi^\dagger_\rightarrow(\mathbf r)\\
\psi^\dagger_\leftarrow(\mathbf r)
\end{array}
\right)
&= \sum_k 
(\gamma_k e^{i k \cdot  r } + \gamma_k^\dagger e^{-i k\cdot r }) 
e^{\kappa z} \\
&\times
\left(
\begin{array}{c}
\sin(k_F z)\\
\sin(k_F z+\theta)
\end{array}
\right)_\sigma
\otimes
\left(
\begin{array}{c}
\cos(\frac{\theta_k + \pi/2}{2})\\
-i \sin(\frac{\theta_k + \pi/2}{2})\\
i \cos(\frac{\theta_k + \pi/2}{2})\\
-\sin(\frac{\theta_k + \pi/2}{2})
\end{array}
\right)_\tau,\nonumber
\end{align}
where $\tan\theta_k = k_y/k_x$. The mode expansion satisfies the Majorana-like conditions: $
\psi_\rightarrow(\mathbf r) = - i \psi^\dagger_\rightarrow(\mathbf r)$ and $\psi_\leftarrow(\mathbf r) =  i 
\psi^\dagger_\leftarrow(\mathbf r)$. For convenience, we define the Majorana operators $\gamma_\alpha(\mathbf r) = 
e^{i\alpha\pi/4} \psi_\alpha (\mathbf r)$ with $\gamma_\alpha^\dagger (\mathbf r) = \gamma_\alpha (\mathbf r)$ for $\alpha = 
\rightarrow,\leftarrow$ and the Majorana operators $\{ \gamma_\rightarrow,\gamma_\leftarrow\}$ transform to $\{\gamma_\leftarrow, -\gamma_\rightarrow\}$ under the TR transformation.

\begin{figure}[tb]
\centerline{
\includegraphics[width=0.7\linewidth]{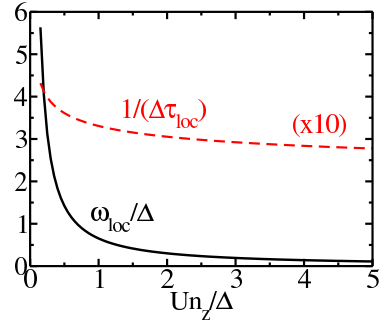} }
\caption{\label{vel} The resonant energy and the lifetime of the localized bound state by varying the coupling strength $U n_z$. The black solid line is the resonant energy $\hbar\omega_{\text loc}/\Delta$ while the red dash line is the relaxation rate ($\hbar/\Delta\tau_{loc}$) of the associated localized state.}
\end{figure} 

The effective Hamiltonian can be rewritten in term of a spinless fermion $f = (\gamma_\rightarrow + 
i\gamma_\leftarrow)/\sqrt{2}$,
\begin{equation}
H_s = v_s\int d^2 \mathbf r \Psi^\dagger(\mathbf r) (i\partial_x 
\tau_y + i\partial_y \tau_x ) \Psi(\mathbf r) ,
\end{equation}
where $\Psi(\mathbf r) = (f(\mathbf r),f^\dagger(\mathbf r))^T$ and $\mathbf \tau = (\tau_x, \tau_y, \tau_z)$ are SU(2) Pauli 
matrices in the Nambu space. Accordingly, the physical spin SU(2) matrices $\mathbf s$ can transform into the new basis, i.e. 
$(s_x,s_y,s_z) \rightarrow (\tau_x,-\tau_y,-\tau_z)$, respectively. 

Because of the Majorana nature of the surface modes, neither the local density operator $\rho(\mathbf r) = 
\sum_\alpha \psi^\dagger_\alpha(\mathbf r)\psi_\alpha(\mathbf r)$ nor the components of the local spin density operators parallel 
to the surface, $\hat s_x = \psi^\dagger_\rightarrow \psi_\rightarrow - \psi^\dagger_\leftarrow \psi_\leftarrow $ and $\hat s_y 
= i \psi^\dagger_\rightarrow \psi_\leftarrow - i \psi^\dagger_\leftarrow \psi_\rightarrow$ can be non-trivial. However it is possible to construct 
the non-trivial component of the spin density operator which is perpendicular to the surface, $\hat s_z = \psi^\dagger_\rightarrow \psi_\leftarrow + 
\psi^\dagger_\leftarrow \psi_\rightarrow = - 2 i \gamma_\rightarrow \gamma_\leftarrow$. The Ising-type spin density is crucial for Majorana boundary modes. Its strong anisotropy reflects the spin-triple pairing symmetry and the spin-orbital coupling in the bulk TSC. In the rest of this paper, we will analyse the effects of magnetic impurities coupled with this gapless Ising spin density. 

\emph{\bf Magnetic impurity induced in-gap state}--- We first consider a static magnetic impurity with large magnetic moment $S$ which scatters the Majorana boundary modes classically with 
interaction strength $J$ at the origin $\mathbf r = 0$. The scattering process can be described by the perturbation $V = \int d\mathbf r \Psi^\dagger(\mathbf r) \langle \mathbf r_1 | \hat V | \mathbf r_2 \rangle \Psi(\mathbf r_2)$ where
\begin{equation}
\label{potential}
\langle \mathbf r_1 | \hat V | \mathbf r_2 \rangle = - U n_z \tau_z \delta (\mathbf r_1) \delta(\mathbf r_2),
\end{equation}
where $U=JS$. Here we have assumed the classical impurity spin $\mathbf S = S \mathbf n$, whose direction is given by the fixed unit vector $\mathbf n = 
(n_x,n_y,n_z)$. In contrast, the spin quantization axis of a quantum impurity spin is determined by the Ising-spin orientation of the surface modes.

The effect of the magnetic impurity scattering can be addressed using the T-matrix technique with the T-matrix,\cite{biswas}
\begin{align} 
\hat T(\omega) 
= \hat V + \hat V \hat G^0_{\rm ret}(\omega) \hat T(\omega)
= \frac{1}{1-\hat V \hat G^0_{\rm ret}(\omega)}  \hat V,
\end{align} 
where $G^0(\mathbf r_1, \mathbf r_2,\omega) = \langle \mathbf r_1 | G^0_{\rm ret}(\omega) | \mathbf r_2\rangle$ is the bare Green function of the effective Hamiltonian $H_s$ in real space,
\begin{align}
\label{G0}
G^0(\mathbf R, \mathbf 0, \omega) &= \frac{i\omega}{4} \left( f_0(R,\omega)I 
\right.\nonumber\\
&\left. + f_1(R,\omega) ( \tau_y\cos\theta_R + \tau_x \sin\theta_R) 
\right).
\end{align}
Here $\theta_R$ is the angle of vector $\mathbf R$ from the x-axis.  $f_0(R,\omega) = -{\rm sgn}(\omega)J_0(|\omega|R) - i Y_0(|\omega|R)$ and 
$f_1(R,\omega) = -i J_1(|\omega| R ) + {\rm sgn}(\omega) Y_1(|\omega| R)$ where $J_i(x)$ , $Y_i(x)$ are the Bessel functions of the first and second kind respectively.

Because of the local form of the scattering potential Eq.\eqref{potential} and the unperturbed on-site Green function $G(\mathbf 0,\mathbf 0,\omega) = g_0(\omega) I$ is diagonal in $\tau_z$, the T-matrix can be evaluated analytically and $T(\mathbf r_1,\mathbf r_2,\omega) = \langle \mathbf r_1 | \hat T(\omega) | \mathbf r_2\rangle$ becomes,
\begin{align}
T(\mathbf r_1,\mathbf r_2,\omega) = \frac{U n_z}{1 - U^2 n_z^2 
g_0^2} ( - \tau_z + U n_z g_0 I)  
\delta(\mathbf r_1) \delta(\mathbf r_2).
\end{align}
Here the unperturbed on-site Green function $g_0(\omega)$ is regulated by a short-distance cutoff  $a_0 \ll a\approx \hbar v_s/\Delta$.

The full Green's function $G_{\rm ret} = G_{\rm ret}^0 + G^0_{\rm ret} T G^0_{\rm ret} = G_{\rm ret}^0 + \delta G_{\rm ret}$ can be computed using the T-matrix and the additional part $\delta G_{\rm ret}$ becomes,
\begin{align}
\label{dG}
\delta G_{\rm ret}(\mathbf r,\mathbf r,\omega) &= B(\omega) \left( 
2i f_0 f_1 (\cos\theta_r \tau_x - \sin\theta_r \tau_y )\right.\nonumber\\
&\left. + (f_0^2 + f_1^2 )\tau_z 
- (f_0^2-f_1^2) U n_z g_0 I\right),
\end{align}
where $f_i = f_i(r,\omega)$ for $i=0,1$ and $B(\omega) = U n_z \omega^2/16\pi(1-U^2 n_z^2 g_0(\omega)^2)$. The pole of $B(\omega)$ determines the position of a localized bound state induced by a magnetic impurity where the imaginary part of its denominator determines the relaxation rate ($1/\tau_{loc}$) of it. As shown in Fig.\ref{vel}, we find that there are always two localized states with energy $\pm \omega_{loc}$ for large enough coupling strength $U n_z$,  and the resonant energy $\omega_{loc}$ goes to zero as $U n_z \rightarrow \infty$. On the other hand, the resonant energy will reach the cutoff value $W$ for small enough value of $U n_z$, which implies that the magnetic impurity cannot induce any localized in-gap states if its magnetic moment pointing close to the material boundary or the coupling strength $U$ to the surface modes is sufficiently weak. 

\begin{figure}[tb]
\begin{tabular}{cc}
\includegraphics[width=0.5\linewidth]{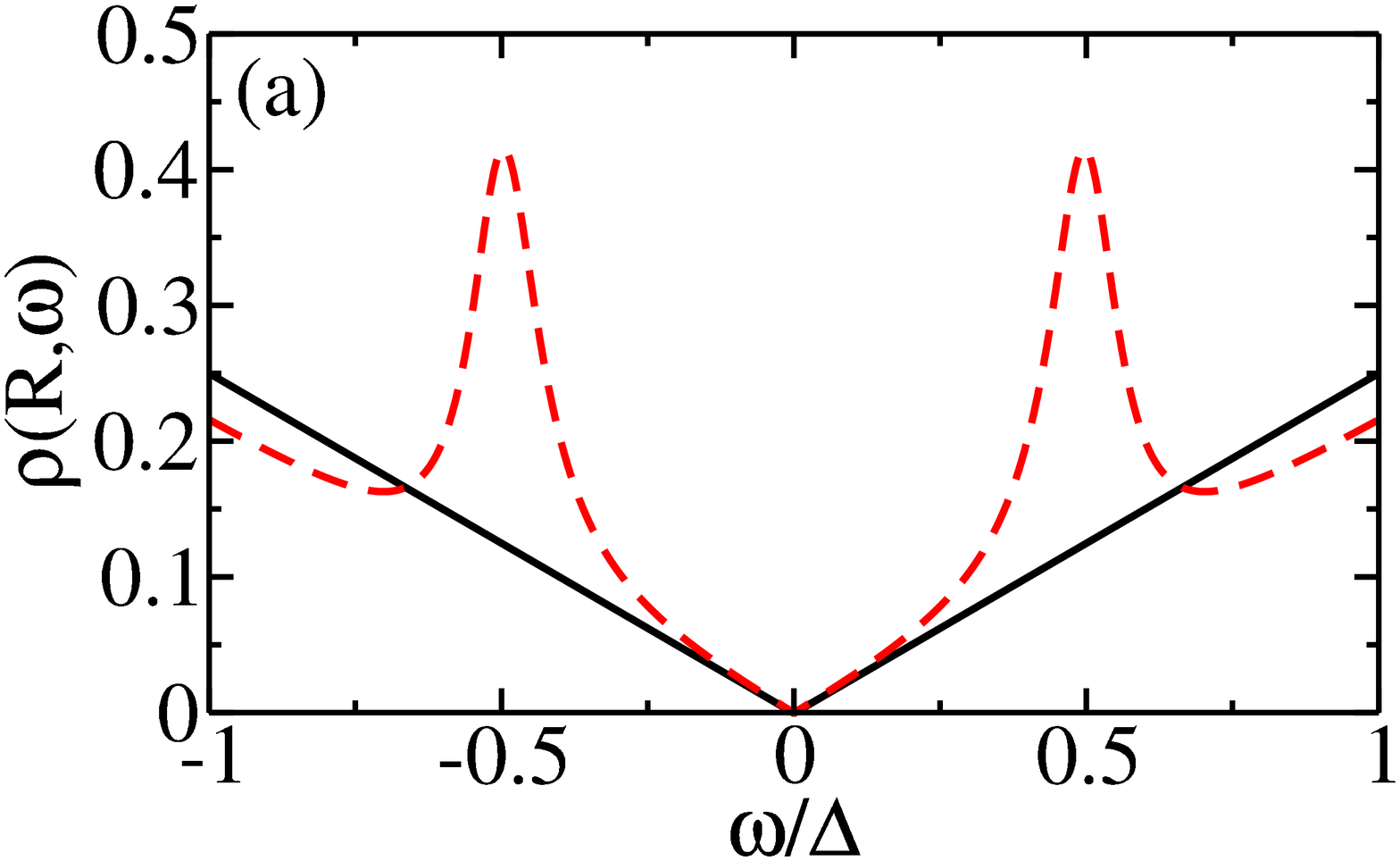} &
\includegraphics[width=0.5\linewidth]{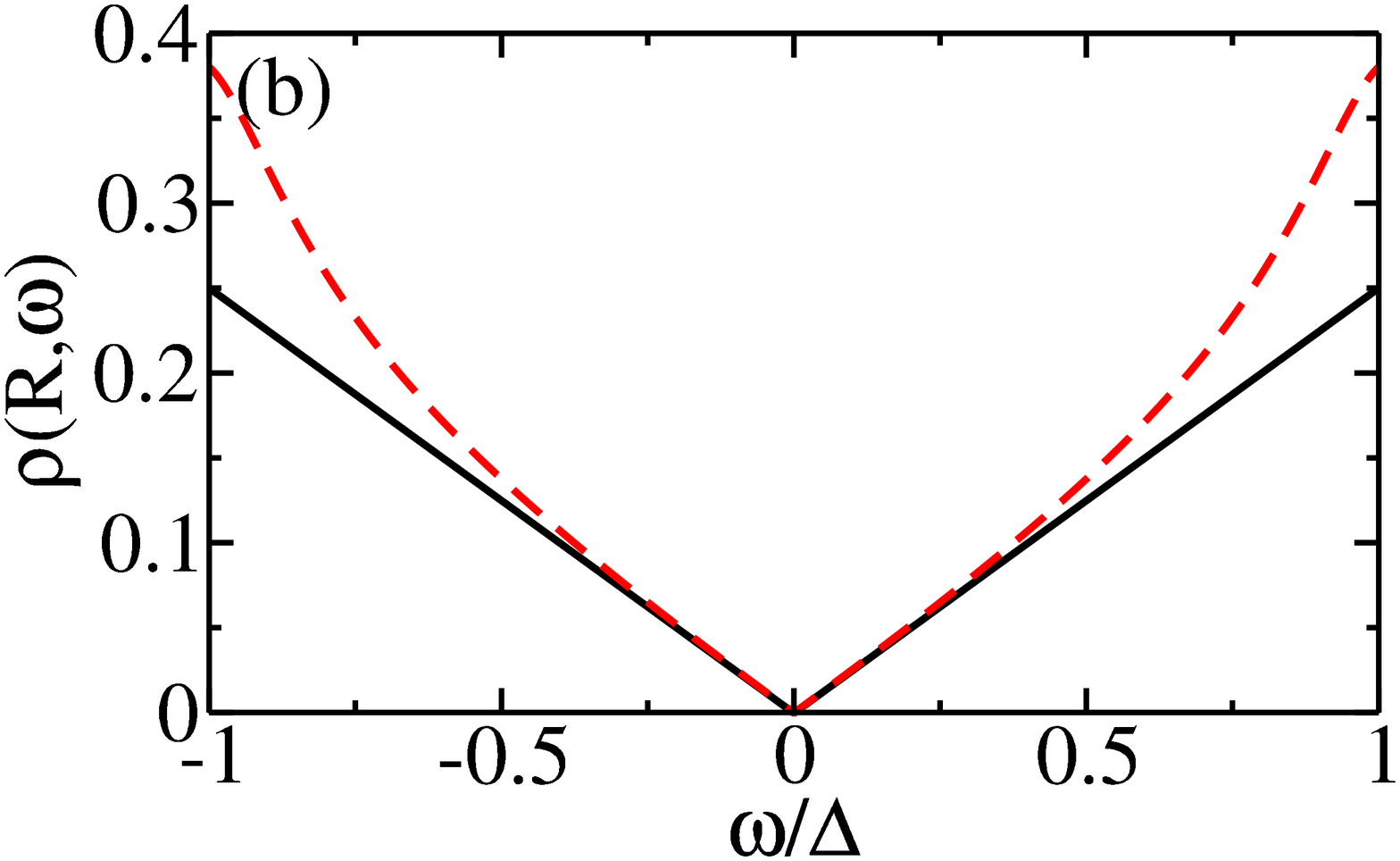}\\
\includegraphics[width=0.5\linewidth]{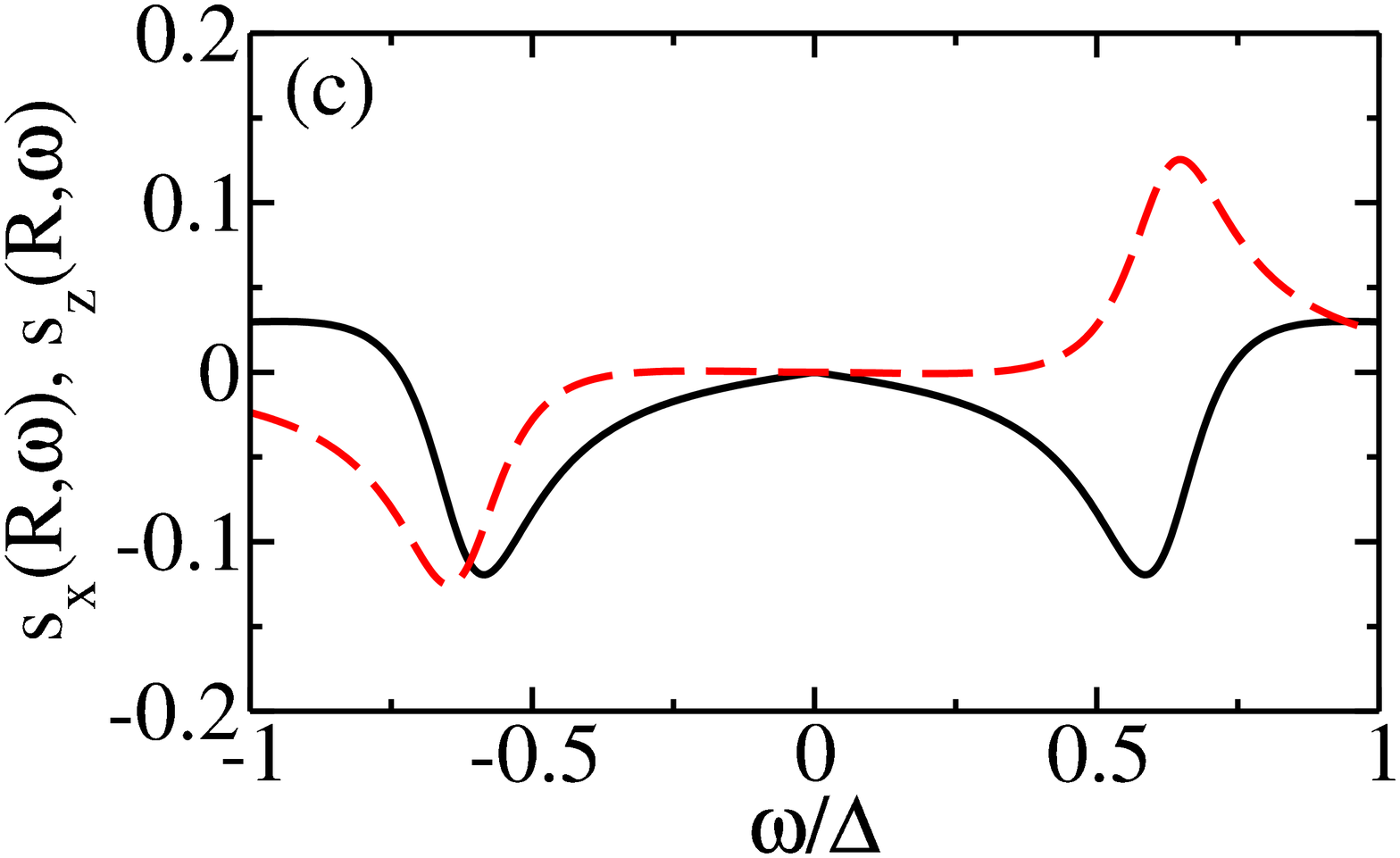} &
\includegraphics[width=0.5\linewidth]{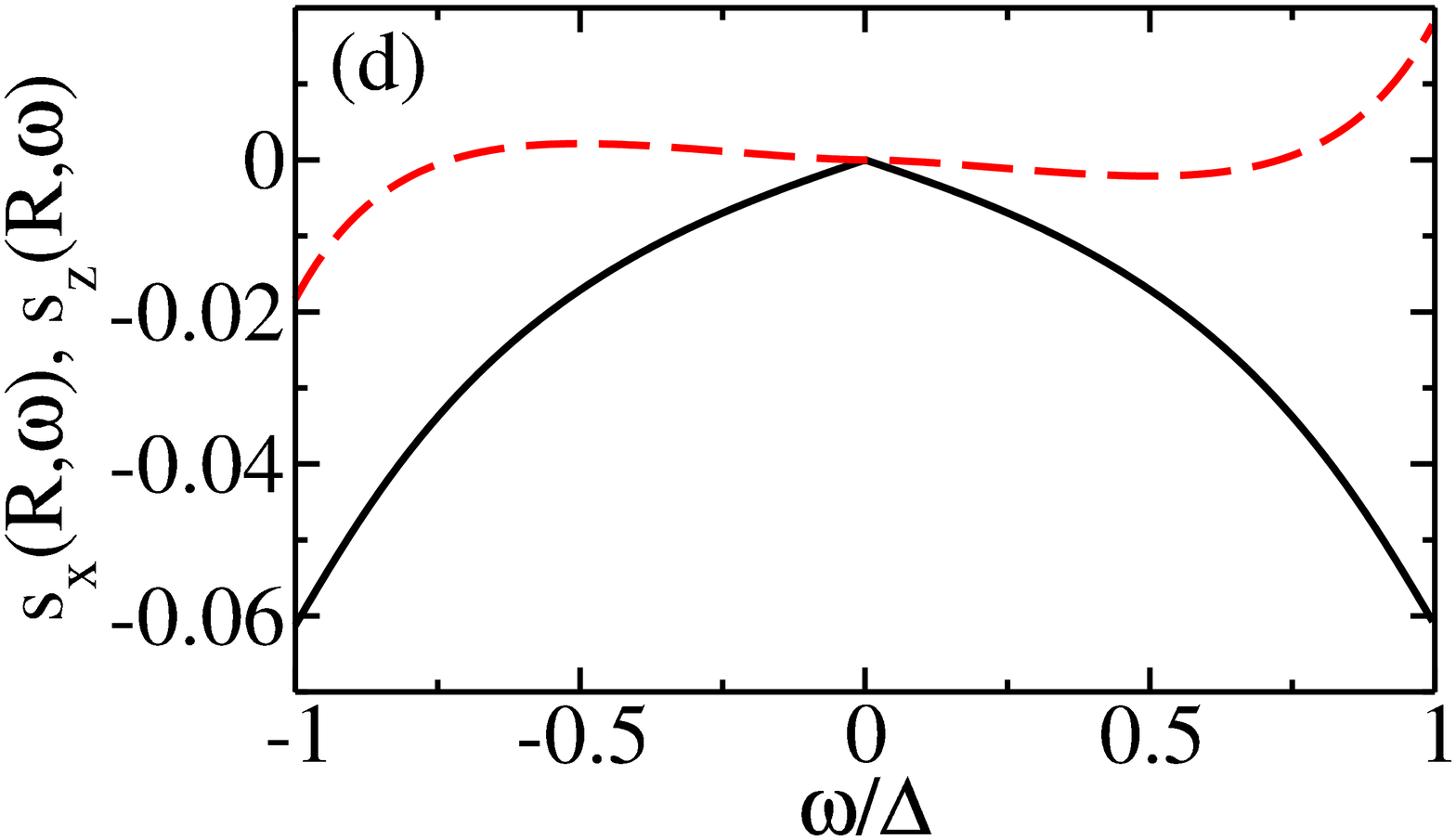}
\end{tabular}
\caption{\label{LDOS}LDOS plots showing the low-energy resonances at $R=0.5a$ away from a  magnetic impurity with (a) $U n_z/\Delta = 1.0$ and (b) $U n_z/\Delta = 0.5$ respectively. Note that the resonance peak is sharper and stronger for larger $U n_z$ and the modulation of the LDOS at $n_z=0.5$ is pronounced at the energy scale beyond the high-energy cutoff $\Delta$ . Energy resolved spin density plot at $\mathbf R=(0.5a,0)$ at scattering strength (c) $Un_z/\Delta=1$ and (d) $Un_z/\Delta=0.5$ respectively.  The black solid line is $s_x(\mathbf R,\omega)$ while the red dash line is $s_z(\mathbf R,\omega)$. We note that the modulation of the spin density is much weaker and extended beyond the high-energy cutoff $\Delta$ at $Un_z = 0.5$. In both cases, $s_y(\mathbf R,\omega)=0$.} 
\end{figure}

The local density of state (LDOS) $\rho(\mathbf R,\omega) = -\frac{1}{\pi} \Im{\rm Tr} G_{\rm ret}(\mathbf 
R,\mathbf R,\omega)$ varies due to the scattering with the magnetic impurity,
\begin{align}
\rho(\mathbf R,\omega) =  \frac{|\omega|}{4} + 2 \Im B(\omega) U n_z (f_0^2-f_1^2).
\end{align}
The linear density of states reflects the linear energy dispersion of the Majorana excitations in the absence of magnetic impurities. As shown in Fig.(\ref{LDOS}a,b), we find low-energy resonances in LDOS induced by a magnetic impurity. The resonance peaks become sharper and approach to zero energy with increasing impurity strength, and they decay away from the impurity as $1/R^2$ which can be determined by the scaling dimension of the Majorana operators. In contrast to the surface of three-dimensional topological insulator where in-gap states can be induced independent  of the impurity moment orientation\cite{biswas}, the modulation of the LDOS $\delta\rho(\mathbf R,\omega)$ shown in Fig.(\ref{LDOS}a,b) is only sensitive to its projection along the z-axis. 

Similarly, the energy-resolved spin density averages, $\mathbf s(\mathbf R,\omega) = -\frac{1}{\pi} \Im {\rm Tr} G_{\rm ret}(\mathbf 
R,\mathbf R,\omega) \frac{\mathbf\sigma}{2}$, which can be measured by the recently developed spin-resolved STM technique\cite{meier}, is found to be
\begin{align}
\label{ss}
\mathbf s(\mathbf R,\omega) =  \Im B(\omega) \left((f_0^2 + f_1^2 )\hat z -2if_0 f_1 (\cos\theta_R \hat x + \sin\theta_R \hat y ) 
\right).
\end{align}
In Fig.(\ref{setup}b,c), we show the energy-resolved spin textures around a magnetic impurity which is pointing in the normal direction. We see that it induces not only the z-direction spin LDOS, but also an in-plane one which is originating from the helical nature of the surface modes. If the magnetic impurity pointing in other directions, the induced magnetization is similar to Fig.(\ref{setup}b,c) qualitatively. However, the magnitude of the induced magnetization linearly proportional to the magnetic spin projection to the Majorana Ising spin direction $n_{z}$. We notice that the out-of-plane energy-resolved spin density is an odd function of energy while the in-plane one is even as shown in Fig.(\ref{LDOS}c,d), which are originated from the PH symmetry on the boundary modes, $J (H_s+V) J^{-1} = - (H_s+V)$ where $J = \tau_x K$ is the effective PH symmetry transformation on the surface. Furthermore, the in-plane energy-resolved spin densities satisfy the sum rules $\int_{-\infty}^{0} s_{x,y}(\mathbf R,\omega)d\omega =0 $ which are also emerged from the Majorana nature of the underlying surface excitations.

\emph{\bf RKKY interaction}---Finally, we consider the dynamics of magnetic impurities interacting with helical Majorana excitations. In particular, we analyse the RKKY interactions between magnetic impurities mediated via the Majorana surface modes. In the DIII TSCs, we have shown that the coupling between the surface helical Majorana states and the magnetic impurities is effectively Ising for $T\ll\Delta$,
\begin{equation}
H_{\rm ex}  = - JS \hat n_z \Psi^\dagger(0) \tau_z \Psi(0),
\end{equation}
and $\hat S_z$ is the spin operator of a magnetic impurity which is projected perpendicular to the surface. The RKKY interaction between two magnetic impurities can be evaluated by integrating out the Majorana modes using their real-space Green function,
\begin{align}
\label{rkky}
H_{\rm RKKY} = J(\mathbf r_1 - \mathbf r_2) n_z(\mathbf r_1) n_z(\mathbf r_2),
\end{align}
where $J(\mathbf R) = J^2 \chi_{zz}(\mathbf R)/4$ and $\chi_{zz}(\mathbf R) = -\frac{2}{\pi}\int_{-\infty}^0 d\omega \Im {\rm 
Tr} [G^0(\mathbf R,\omega)\tau_z G^0(-\mathbf R,\omega) \tau_z]$ is the spin susceptibility of the Majorana 
modes, which can be evaluated analytically for 2D helical Majorana fermions
\begin{align}
\label{xzz}
\chi_{zz}(\mathbf R) = -\frac{a^4}{8\pi \hbar |v_s|  R^3}.
\end{align}
In evaluating the spin susceptibility $\chi_{zz}$, a soft cutoff function $e^{-\omega/\omega_0}$ is used and we take the limit $\omega_0\rightarrow\infty$ after performing the integrals\cite{saremi}. The RKKY interactions can also be derived from the energy-resolved spin average, Eq.\eqref{ss}, if we integrate all the filled states with energy up to the chemical potential, $\mu = 0$.

From Eq.\ref{xzz}. We see that the Ising interactions between two impurities are always ferromagnetic. Interactions of an ensemble of magnetic impurities give rise to an ordered ferromagnetic phase with spins pointing perpendicular to the surface and breaks the TR symmetry spontaneously, driving the surface state into a gapped one. According to Eq.\eqref{xzz}, the ordering temperature can be estimated as $k_B T_c\approx J^2 a^4 n_{\rm imp}^{3/2} /\hbar v_s$, where $n_{\rm imp}$ is the impurity density.  At the mean-field level, the ordering magnetic impurities open a mass gap $m = n_{\rm imp} J S$. Such a mass term breaks the TR symmetry of the surface modes and exhibits an anomalous quantum Hall effect, provides a half quantized Hall thermal conductance $\sigma_H =  {\rm sgn}(m)/2h$.\cite{abamin,liu} It is in contrast to the effects of dense magnetic impurities on the surface of three dimensional TI where the RKKY interactions between magnetic impurities mediated via the helical fermionic surface modes are frustrated, and result in a disorder spin-glass phase in which the TR symmetry is still preserved.\cite{abamin}


\emph{\bf Conclusion}--- In summary, we find that the local magnetic impurity can induced a pair of localized in-gap states on the 2D surface of three-dimensional DIII TSCs. Importantly, the energy of the induced in-gap states is sensitive to the orientations between the magnetic impurity and the MF Ising spin direction. We also show that the RKKY-like interactions between magnetic impurities are ferromagnetic Ising. Therefore, at large densities of magnetic impurities, long-range magnetic ordering is developed and  TR symmetry on the surface is broken spontaneously. 

\emph{\bf Acknowledgement}--- We thank the discussion with X. J. Liu and R. Shindou. This work is supported by HKRGC through Grant 605512 and HKUST3/CRF09.

\end{document}